\documentclass[pra,twocolumn,floatfix,aps,superscriptaddress,amsfonts]{revtex4}
\usepackage{graphicx}
\usepackage{array}
\usepackage{amsthm}
\usepackage{amsmath}
\usepackage{amssymb}
\usepackage{multirow}
\newtheorem{Th}{ Theorem}
\newtheorem{Def}{ Definition}
\newtheorem{Ex}{ Example}

\newtheorem{Lem}[Th]{ Lemma}

\newtheorem{Cor}[Th]{ Corollary}

\newcommand{\p}[1]{{ Proof.} #1 \ $\blacksquare$}
\begin{document}
\title{Quantum Access Structure and Secret Sharing}%
\author{Chen-Ming Bai}
\affiliation{
College of Mathematics and Information Science, Shaanxi Normal University, Xi'an 710119, China}

\author{Zhi-Hui Li}%
\email{lizhihui@snnu.edu.cn}
\affiliation{
College of Mathematics and Information Science, Shaanxi Normal University, Xi'an 710119, China}

\author{Yong-Ming Li}
\email{liyongm@snnu.edu.cn}%
\affiliation{College of Computer Science, Shaanxi Normal University, Xi'an 710119, China}
%

\begin{abstract}
In this paper we define a kind of decomposition for a quantum access structure. We propose a conception of minimal maximal quantum access structure and obtain a sufficient and necessary condition for minimal maximal quantum access structure, which shows the relationship between the number of minimal authorized sets and that of the players. Moreover, we investigate the construction of efficient quantum secret schemes by using these techniques, a decomposition and minimal maximal quantum access structure. A major advantage of these techniques is that it allows us to construct a method to realize a general quantum access structure. For these quantum access structures, we present two quantum secret schemes via the idea of concatenation or a decomposition of a quantum access structure. As a consequence, the application of these techniques allow us to save more quantum shares and reduce more cost than the existing scheme.
\end{abstract}
\eid{identifier}
\maketitle
\section{Introduction}

Secret sharing, first introduced by Shamir${[1]}$ and Blakley${[2]}$, is an important cryptographic primitive and then extended to the quantum field${[3-6]}$.
The central aim of protocol is for a dealer to distribute a piece of secret information (called the secret) among a finite set of players $\mathcal{P}$ such that only qualified subsets can collaboratively recover the secret. Traditionally both secret and shares were classical information. While the secret in a quantum scheme may be either an unknown quantum state or a classical one. In the quantum scenario all players are comprised of quantum systems, and they can utilize the quantum communication technique. Compared to the classical secret sharing, quantum secret sharing (QSS) is more secure due to the application of quantum communication technique.
In 1999, Hillery {\it et al.} ${[3]}$ firstly proposed a protocol of QSS by using GHZ states where an unknown qubit can be shared with two players such that to recover the original qubit the players have to put their pieces of quantum information together. Cleve, Gottesman and Lo ${[4]}$ presented a more general scheme. In 2004, Xiao {\it et al.} ${[7]}$ generalized the QSS of Hillery {\it et al.} into arbitrary multiparty. From then on, with the development of quantum cryptography that is unconditional secure in theory, QSS has attracted much attention and progressed quickly in recently years ${[8-24]}$ (for an incomplete list).

The access structure of a secret sharing scheme is a family of all authorized sets. In a classical secret sharing scheme, some researchers have proposed many interesting results${[25,26,27]}$. In quantum case there are also many nice results. For example, Cleve {\it et al.} ${[4]}$ proposed an efficient construction of all threshold schemes and introduced the quantum access structure. Adam Smith ${[28]}$ researched the quantum access structure in detail and used the monotone span programs to design a quantum secret sharing. Marin {\it et al.} ${[29]}$ gave graphical characterisation of the access structure to both classical and quantum information. Gheorghiu ${[14]}$ provided a systematic way of determining the access structure. In Ref.${[5]}$ Gottesman systematically presented a variety of results on the theory of QSS and also defined a maximal quantum access structure. This access structure has some special properties. For example the authorized and unauthorized sets are complements of each other. It plays an important role for  these properties to construct the secret scheme. Moreover,
the maximal quantum access structure also has a very close relationship with pure state quantum secret sharing scheme that encode pure state secrets as pure states (when all of the shares are available). Gottesman also showed the fact there is always a pure quantum secret sharing scheme to realize a maximal quantum access structure. However, in that reference, Gottesman didn't give a discussion about it in detail.
In this paper we further analyze the maximal access structure and give a formal definition. After analyzing the above access structure, we present a minimal maximal quantum access structure, which the number of the minimal authorized sets cannot be reduced when the number of participants is unchanged. We also obtain a sufficient and necessary condition for minimal maximal quantum access structure, which shows the relationship between the number of minimal authorized sets and that of the players.
After analyzing  the minimal maximal access structure is more compact and easier to obtain than the maximal one.

On the other hand, Gottesman combined the maximal access structure with the original access structure and proposed a quantum secret sharing protocol for a general access structure by a threshold cascade scheme. If $\mathcal{S}_1$ and $\mathcal{S}_2$ are quantum secret sharing schemes, then the scheme formed by concatenating them (expanding each share of $\mathcal{S}_1$ as the secret of $\mathcal{S}_2$ ) is also secret sharing. This idea is very good and interesting. However, in his scheme, there are two disadvantages. One is complex to select threshold schemes according to the number of minimal authorized sets. So this will lead to require more quantum resources in this scheme.
As quantum data is expansive and hard to deal with, it would be desirable to use as little quantum data as possible in order to share a secret.
Another is based on the maximal quantum access structure because it is complicated. In the process of testing the real authorized sets, there will increase a lot of work because the number of minimal authorized sets is uncertain. Therefore, it will lead to reduce the efficiency of the scheme.
In this paper we define a decomposition of the quantum access structure to solve the first problem. In these decompositions, we can find an optimal one and use it to reduce the amount of quantum data. For the second problem, we replace the maximal quantum access structure with the minimal maximal quantum access structure. We can easily obtain this minimal maximal quantum access structures, and each minimal maximal quantum access structure includes maximal one. By combining the optimal decomposition and the minimal maximal quantum access structure, we improve Gottesman's scheme and present a more convenient solution than before.
For the optimal decomposition, we also propose a quantum secret sharing scheme to realize a general access structure and compare these schemes.

The structure of the paper is organized as follows. In Sec.II, we define a decomposition of a quantum access structure, explore the maximal quantum access structure and propose a minimal maximal quantum access structure. We also show some results about minimal maximal quantum access structure.
In Sec.III, we propose two schemes realizing the general access structure. One uses a decomposition of quantum access structure, and another is based on the optimal decomposition and the cascade method. The paper is ended with the conclusion and discussion in Sec.VI.
\section{Quantum Access structure}
\subsection{Decomposition of Quantum Access structure}
Quantum access structure plays an important role in quantum secret sharing, and let us give its definition.

\begin{Def}
\rm Let $\mathcal{P}$ be a set of players, the access structure of a secret sharing is the family of authorized sets, $\Gamma\subseteq2^\mathcal{P}$. $\Gamma$ is called a
quantum access structure on $\mathcal{P}$ if it satisfies that

  (a) If $A\subseteq B$ for any $A$ in $\Gamma$, then $B\in\Gamma$;

  (b) If $A, B\in\Gamma$, then $A\cap B\neq \emptyset$.
\end{Def}
By Definition 1, it is obvious that a quantum access structure
must satisfy the monotonicity and the no-cloning theorem ${[30,31]}$. For convenience, in the following $\mathcal{P}$ shows the set of participants and $\Gamma$ represents a access structure on $\mathcal{P}$.

In the classical secret sharing, many researchers have proposed a decomposition of an access structure ${[26,32]}$. Similarly we present a decomposition of a quantum access structure and later will use this decomposition to realize a general access structure in Sec.III.A.

\begin{Def}
\rm Given a quantum access structure $\Gamma$ containing $r$ minimal authorized sets, a decomposition of $\Gamma$ is composed by a set $\{\Gamma_1, \Gamma_2,\cdots,\Gamma_l\}$, where $\Gamma_i \ (i=1,2,\cdots,l)$ satisfies the following conditions:

(a) $\Gamma_i\subseteq \Gamma$ and $\Gamma=\bigcup_{i=1}^{l}\Gamma_i$;

(b) $\Gamma_i\cap\Gamma_j =\emptyset$ for any $\Gamma_i,\Gamma_j\subseteq \Gamma\ (i\neq j)$;

(c) There exists quantum secret sharing protocol realizing a quantum access structure $\Gamma_i \ (i=1,2,\cdots,l)$. Furthermore, if $l=r$, then the decomposition is trivial;
if $l< r$, then the decomposition is called $l$-decomposition. For an $l$-decomposition, if there doesn't exist a positive integer $l'$ such that $l'<l$, then this decomposition is optimal.
\end{Def}

\textbf{Remark:}
A decomposition of the access structure is defined based on the number of partition for the quantum access structure $\Gamma$.
Because the partition of $\Gamma$ is not unique, the decomposition is not unique.

Suppose that $\Gamma$ is a quantum access structure and $\{\Gamma_1, \Gamma_2,\cdots,\Gamma_l\}$ is a decomposition of $\Gamma$. When $\Gamma_i=\{A_{i1},A_{i2},\cdots,A_{ir_i}\}\ (i=1,2,\cdots,l)$, then $\Gamma$ is denoted by
$\Gamma=\{A_{11},A_{12},\cdots,A_{1r_1},\cdots,A_{l1},A_{l2},\cdots,A_{lr_l}\}.$

\subsection{Minimal Maximal Quantum Access structure}
In Ref.${[5]}$, Gottesman introduced a maximal quantum access structure in which the authorized and unauthorized sets are complement of each other. In the following let us formally define a maximal quantum access structure.

\begin{Def}
\rm Let $\mathcal{P}$ be a set of players, $\Gamma$ a quantum access structure and $ \mathcal{A}$ a set of all unauthorized groups. Then $\Gamma$ is said to be a maximal quantum access structure, denoted by $\Gamma_M$, if it satisfies that

(a) If any $A\in\Gamma$, then $\overline{A}\in\mathcal{A}$;

(b) If any $B\in\mathcal{A}$, then $\overline{B}\in \Gamma$.

where $\mathcal{A}=\{A\in 2^\mathcal{P}|A\notin\Gamma \ \rm{and}\ A\neq{\emptyset}\}$,
$\overline{A}=\mathcal{P}\setminus A$ and $\overline{B}=\mathcal{P}\setminus B$.
\end{Def}

By Definition 3, we can know that if $A\subsetneqq B$ for any minimal authorized set $B$ in $\Gamma_M$, the complement of the set $A$ must be authorized. Next we analyze the determination and properties of a maximal quantum access structure. Firstly, we give the following lemma.

\begin{Lem}
 \rm([33])  Let $\Gamma\subseteq2^\mathcal{P}$ be a general quantum access structure and $ \mathcal{A}= \mathcal{A}_{1} \cup\mathcal{A}_{2}$ a set of all unauthorized groups, where $\mathcal{A}_{1}=\{A\in\mathcal{A}\ |\ \exists\ B\in\Gamma, A\cap B=\emptyset \}$ and $\mathcal{A}_{2}=\{A\in\mathcal{A}\ |\ \forall\ B\in\Gamma, A\cap B\neq\emptyset \}$.

(i) If $A\in \mathcal{A}_{1}$, then $\overline{A}\in\Gamma$.

(ii) If $A\in \mathcal{A}_{2}$, then $\overline{A}\in\mathcal{A}_{2}\subseteq\mathcal{A}$.
\end{Lem}

\begin{Th}
 \rm Let $\Gamma$ be a quantum access structure and $ \mathcal{A}= \mathcal{A}_{1} \cup\mathcal{A}_{2}$ a set of all unauthorized sets, where $\mathcal{A}_{1}=\{A\in\mathcal{A}\ |\ \exists\ B\in\Gamma, A\cap B=\emptyset \}$ and $\mathcal{A}_{2}=\{A\in\mathcal{A}\ |\ \forall\ B\in\Gamma, A\cap B\neq\emptyset \}$. Then
$\Gamma$ is a maximal quantum access structure if and only if $\mathcal{A}= \mathcal{A}_{1}$, i.e., $\mathcal{A}_{2}=\emptyset$.
\end{Th}

\p{\rm Suppose that $\mathcal{A}_{2}\neq\emptyset$. By Lemma 1, we can obtain that $\overline{A}\in \mathcal{A}_{2}\subseteq \mathcal{A}$ for any $A$ in $\mathcal{A}_{2}$. Thus both $A$ and $\overline{A}$ are unauthorized sets, and this leads to a contradiction with the maximal quantum access structure $\Gamma$. Therefore $\mathcal{A}_{2}=\emptyset$.

For the converse, we can get that $\mathcal{A}= \mathcal{A}_{1}$ since $\mathcal{A}_{2}=\emptyset$.
By Lemma 1, it implies that $\overline{A}\in \Gamma$ for all $A$ in $\mathcal{A}_{1}$. Since the fact that the quantum access structure satisfies the no-cloning theorem, we can find that $\overline{B}\in \mathcal{A}_{1}=\mathcal{A} $ for all $B$ in $\Gamma $. According to Definition 3, $\Gamma$ must be a maximal quantum access structure.}

\begin{Th}
\rm Let $\mathcal{P}$ be a set of players and $\Gamma$ a quantum access structure on $\mathcal{P}$. There are always some subsets of $\mathcal{P}$ added to $\Gamma$ such that $\Gamma$ becomes a maximal quantum access structure.
\end{Th}

\p{\rm Let $\mathcal{P}$ be a set of players, $\Gamma\subseteq 2^\mathcal{P}$ a quantum access structure and $\mathcal{A}$ a set of all unauthorized groups. Suppose that $\Gamma$ can be denoted by $\Gamma=\{A_1,A_2,\cdots,A_r\}$, where $A_i\in2^\mathcal{P}\ (i=1,2,\cdots,r)$ is the minimal authorized set.
If $\Gamma$ is a maximal quantum access structure, this proposition is obviously true.
If $\Gamma$ isn't a maximal quantum access structure, we can construct the maximal access structure. Since that $\Gamma$ isn't maximal, we can find that all sets $B_1,B_2,\cdots,B_m$ are in $\mathcal{A}$ and the complements of them, $\overline{B}_1,\overline{B}_2,\cdots,\overline{B}_m$, are also in $\mathcal{A}$. For convenience, $\mathcal{S}$ represents a set $\{B_1,B_2,\cdots,B_m,\overline{B}_1,\overline{B}_2,\cdots,\overline{B}_m\}$.
Adding the set $C_{j_1}\in\mathcal{S}$ to the access structure $\Gamma$, then we can obtain a new quantum access structure $\Gamma'=\{A_1,A_2,\cdots,A_r,C_{j_1}\}$.
Continuing to add the set $C_{j_2}\in \mathcal{S}$ to $\Gamma'$, where $C_{j_2}$ should satisfy the conditions: $C_{j_1}\nsubseteq C_{j_2}$ and $C_{j_2}\cap C_{j_1}\neq\emptyset$, so we can have another access structure $\Gamma''=\{A_1,A_2,\cdots,A_r,C_{j_1},C_{j_2}\}$. Repeat the above process
until there doesn't exist sets meeting the conditions. Since $2^\mathcal{P}$ is finite, we can obtain the maximal quantum access structure.}

Theorem 3 tells us that it can get a maximal quantum access structure for any quantum access structure and show that how we construct a maximal quantum access structure through a quantum access structure. In order to understand we provide an example of an access structure on the set $\mathcal{P}$ with five players.

\begin{Ex}
\rm Given the set of players $\mathcal{P}=\{P_1,P_2,P_3,P_4,P_5\}$ and the quantum access structure $\Gamma=\{P_1P_2,P_1P_4P_5,P_2P_3P_5,P_2P_3P_4\}$. For the access structure $\Gamma$, it must satisfy the monotonicity. Therefore it means that these sets containing authorized sets in $\Gamma$ are authorized. Since the no-cloning theorem, the complements of these authorized sets are unauthorized. Apart from the above sets, the remaining sets are denoted by $P_1P_3,P_2P_4,P_2P_5,P_2P_4P_5,P_1P_3P_5,P_1P_3P_4$.

If we add $P_1P_3$ to $\Gamma$, that is, $P_1P_3$ becomes an authorized set, then $P_1P_3P_5$ and $P_1P_3P_4$ are authorized and the others are unauthorized. So we obtain a maximal quantum access structure $$\Gamma_M=\{P_1P_2,P_1P_3,P_1P_4P_5,P_2P_3P_5,P_2P_3P_4\}.$$

If we add $P_2P_4P_5$, $P_1P_3P_5$ and $P_1P_3P_4$ to $\Gamma$, we will obtain another maximal quantum access structure
\begin{eqnarray*}
\Gamma_M'&=&\{P_1P_2,P_1P_3P_5,P_1P_3P_4,P_1P_4P_5,P_2P_3P_5,\\
&\quad&P_2P_3P_4,P_2P_4P_5\}.
\end{eqnarray*}
\end{Ex}

By this example, we can get different maximal quantum access structures after adding different sets to the same quantum access structure. The number of minimal authorized sets contained in each maximal access structure is not equal. For Example 1, if some authorized sets in $\Gamma_M'$ are changed, for example, two sets $P_1P_3P_4$ and $P_1P_3P_5$ are replaced with $P_1P_3$ and the set $P_2P_4P_5$ is deleted, then we can obtain a new maximal quantum access structure, i.e. $\Gamma_M$. If we continue to change the minimal authorized set in $\Gamma_M$, we will find that the number of participants in the new maximal access structure will be reduced. Based on this fact, we propose the definition of minimal maximal quantum access structure.

\begin{Def}
\rm Let $\mathcal{P}$ be a set of players and $\Gamma_M$ a maximal quantum access structure. $\Gamma_M$ is called a minimal maximal quantum access structure on $\mathcal{P}$, denoted by $\Gamma_M^{(m)}$, if it satisfies that the number of the minimal authorized sets in $\Gamma_M$ cannot be reduced when the number of participants is unchanged.
\end{Def}
It is easy to verify that $\Gamma_M$ is a minimal maximal access structure in Example 1. How do we change the given maximal access structure to a minimal maximal one? The following theorem shows this construction.

\begin{Th}
\rm Let $\Gamma_M$ be a maximal quantum access structure. Then a minimal maximal quantum access structure is given by changing some authorized sets of $\Gamma_M$.
\end{Th}

\p{\rm Let $\Gamma_M$ be a quantum access structure and it can be denoted by $\Gamma_M=\{A_1,A_2,\cdots,A_r\}$, where $A_i\ (i=1,2,\cdots,r)$ is the minimal authorized set. First we can take some minimal authorized sets $A_{j_1},A_{j_2},\cdots,A_{j_k} (k<r)$. Then $\bigcap_{j\in\{j_1,j_2,\cdots,j_k\}}A_j=B_l$ containing at least 2 players. Use $B_l$ instead of $A_j(\supseteq B_l)$ and delete the set $\overline{B}_l$. Hence we can obtain a new maximal quantum access structure. Repeat the above process until the number of minimal authorized sets cannot be reduced, we will be forced to stop. At this time, we obtain a minimal maximal quantum access structure.}

\begin{Ex}
\rm Given the set of players $\mathcal{P}=\{P_1,P_2,P_3,P_4,P_5,P_6\}$ and the maximal quantum access structure is denoted by
\begin{eqnarray*}
\Gamma_M&=&\{P_1P_2,P_1P_3P_4,P_1P_3P_5,P_1P_3P_6,\\
&\quad&P_1P_4P_5,P_1P_4P_6,P_1P_5P_6,P_2P_3P_5P_6,\\
&\quad&P_2P_4P_5P_6,P_2P_3P_4P_5,P_2P_3P_4P_6\}.
\end{eqnarray*}
Without lost of generality, we may take $P_1P_3P_4, P_1P_3P_5$ and $P_1P_3P_6$. Since
$(P_1P_3P_4)\cap(P_1P_3P_5)\cap(P_1P_3P_6)=P_1P_3$, we can replace $P_1P_3P_4$, $P_1P_3P_5$ and $P_1P_3P_6$ with $P_1P_3$ and delete the set $P_2P_4P_5P_6$. Then we obtain the new access structure
\begin{eqnarray*}
\Gamma'&=&\{P_1P_2,P_1P_3,P_1P_4P_5,P_1P_4P_6,P_1P_5P_6,\\
&\quad&P_2P_3P_4P_5,P_2P_3P_4P_6,P_2P_3P_5P_6\}.
\end{eqnarray*}
At the same method, we can also continue to replace
$P_1P_4P_5$ and $P_1P_4P_6$ with $P_1P_4$ and delete the set $P_2P_3P_5P_6$.
Hence we can obtain the new maximal quantum access structure
$$\Gamma''=\{P_1P_2,P_1P_3,P_1P_4,P_1P_5P_6,P_2P_3P_4P_5,P_2P_3P_4P_6\}.$$
If we continue to change the minimal authorized set, some participants will not appear in the new authorized set. Hence $\Gamma'$ is a minimal maximal quantum access structure.
\end{Ex}

This example shows the relationship between the number of minimal authorized sets and that of the participants. Thus we present a sufficient and necessary condition about the minimal maximal quantum access structure. In order to prove the condition, we need give the following lemma.

\begin{Lem}
 \rm Let $\mathcal{P'}=\{P_1,\cdots,P_{n-1}\}$ be a set of the players and $\Gamma_M=\{A_1,A_2,\cdots,A_r\}$ a maximal quantum access structure on $\mathcal{P'}$, where $A_i\ (i=1,2,\cdots,r)$ is the minimal authorized set. If a player $P_n$ is added to $A_i$ in $\Gamma_M$, then the new quantum access structure isn't maximal.
\end{Lem}

\p{\rm Suppose that $\mathcal{P}=\mathcal{P'}\cup\{P_n\}$, then the new quantum access structure on $\mathcal{P}$ can be denoted by $\Gamma'=\{A_1,\cdots,A_{i-1},A_i\cup\{P_n\},A_{i+1}\cdots,A_r\}$, where $A_i\in\Gamma_M (i=1,2,\cdots,r)$. Since that $A_i\cup\{P_n\}$ is a minimal authorized set of $\Gamma'$ and $A_i\subsetneqq A_i\cup\{P_n\}$, then we know that $A_i$ is an unauthorized set.

Next we need to prove that $\mathcal{P}\setminus A_i$ is an unauthorized set, that is, $A_i\cup\{P_n\}\nsubseteq \mathcal{P}\setminus A_i$ and $A_j\nsubseteq\mathcal{P}\setminus A_i (j\neq i)$, where $\mathcal{P}\setminus A_i=\overline{A}_i\cup\{P_n\}$.
If $A_i\cup\{P_n\}\subseteq \overline{A}_i\cup\{P_n\}$, then $A_i\subseteq \overline{A}_i$. Obviously, this leads to a contradiction. If $A_j\subseteq \overline{A}_i\cup\{P_n\}$, then $A_j\subseteq \overline{A}_i$, i.e., $A_j\cap A_i=\emptyset$. This contradicts the fact that $A_j\cap A_i\neq\emptyset$ for any $A_j, A_i$ in $\Gamma_M$. Hence $\mathcal{P}\setminus A_i$ is also an unauthorized set. Both $A_i$ and $\mathcal{P}\setminus A_i$ are unauthorized, so $\Gamma'$ isn't a maximal quantum access structure on $\mathcal{P}$.}

\begin{Th}
\rm Let $\mathcal{P}$ be a set with $n$ players and $\Gamma_M$ a maximal quantum access structure containing $r$ minimal authorized sets. Then $\Gamma_M$ is a minimal maximal quantum access structure if and only if $r=n$.
\end{Th}

\p{\rm
($\Leftarrow$) Since the maximal quantum access structure $\Gamma_M$ contains $r$ minimal authorized sets and $r=n$,
 we can denote $\Gamma_M=\{A_1,A_2,\cdots,A_n\}$. If some minimal authorized sets of $\Gamma_M$ are changed, then we can obtain a new quantum access structures $\Gamma'=\{B_1,B_2,\cdots,B_m\} (m<n)$.
If $\Gamma'$ is not a maximal quantum access, then the theorem is true.
If $\Gamma'$ is a maximal quantum access, then we can find a player $P_{i_0} \notin B_j$ for each $B_j$ in $\Gamma'$. Otherwise there exists a set $B_{j'}$ such that $P_{i_0} \in B_{j'}$. By Lemma 5, we know that $\Gamma'$ is not a maximal quantum access. This leads to a contradiction. Hence $\Gamma_M$ is a minimal maximal quantum access structure.

($\Rightarrow$) For $\mathcal{P}=\{P_1,P_2,P_3\}$, the minimal maximal quantum access structure on $\mathcal{P}$ can be denoted by $\Gamma_M^{(m)}=\{P_1P_2,P_1P_3,P_2P_3\}$. Obviously, the conclusion is true.

For $\mathcal{P}=\{P_1,P_2,P_3,P_4\}$, the minimal maximal quantum access structure on $\mathcal{P}$ can be denoted by $\Gamma_M^{(m)}=\{P_1P_2,P_1P_3,P_1P_4,P_2P_3P_4\}$. It is obvious to see that the conclusion holds.

For $\mathcal{P}=\{P_1,P_2,P_3,P_4,P_5\}$, all minimal maximal quantum access structures on $\mathcal{P}$ can be denoted by
\begin{eqnarray*}
  \Gamma_{M_1}^{(m)}&=&\{P_1P_2,P_1P_3,P_1P_4,P_1P_5,P_2P_3P_4P_5\} \\
  \Gamma_{M_2}^{(m)}&=&\{P_1P_2,P_1P_3,P_1P_4P_5,P_2P_3P_4,P_2P_3P_5\}
\end{eqnarray*}
Obviously, the number of minimal authorized sets in each minimal maximal quantum access structure is equal to that of the players. Hence the conclusion is true.

When there are $n-1$ players, i.e., $\mathcal{P'}=\{P_1,P_2,\cdots,P_{n-1}\}$, the minimal maximal quantum access structure on $\mathcal{P'}$ can be denoted by $\Gamma_M=\{A_1,A_2,\cdots,A_r\}$, where $A_i \ (i=1,2,\cdots,r)$ is the minimal authorized set. We assume that this conclusion is true, that is, $r=n-1$.

Next we need prove that when there are $n$ players, this conclusion is also true.
Suppose that $\mathcal{P}=\mathcal{P'}\cup\{P_n\}$, we add the player $P_n$ to $A_i$ in $\Gamma_M$, where $A_i$ satisfies that for each $B\subsetneqq\overline{A}_i=\mathcal{P'}\setminus A_i$ there exists $A_j (j\neq i)$ in $\Gamma_M$ such that $B\cap A_j=\emptyset$.
Then we can obtain a new quantum access structure $\Gamma'$ and it is denoted by $\Gamma'=\{A_1,\cdots,A_{i-1},A_i\cup\{P_n\},A_{i+1}\cdots,A_r\}$.
By Lemma 5, we know that $\Gamma'$ isn't maximal.
From the proof of Lemma 5 we find the unauthorized sets $A_i$ and $\mathcal{P}\setminus A_i=\overline{A}_i\cup\{P_n\}$. Add $\overline{A}_i\cup\{P_n\}$ to $\Gamma'$ and obtain a quantum access structure $$\Gamma''=\{A_1,\cdots,A_{i-1},A_i\cup\{P_n\},A_{i+1}\cdots,A_r,\overline{A}_i\cup\{P_n\}\}$$
It is easy to verify that $\Gamma''$ is a maximal quantum access structure. Without loss of generality, in the following we may take $A_i\cup\{P_n\}\in \Gamma''$ as an example, and the others can be analyzed by the same method.

Case 1: Since $A_i\subsetneqq A_i\cup\{P_n\}$, the set $A_i$ is unauthorized. The complement of $A_i$ is $\mathcal{P}\setminus A_i=\overline{A}_i\cup\{P_n\}$ and it is an authorized set, so this case holds.

Case 2: If $B\subsetneqq A_i$, then $B\cup\{P_n\}\subsetneqq A_i\cup\{P_n\}$. So $B\cup\{P_n\}$ is an unauthorized set. The complement of $B\cup\{P_n\}$ is $\mathcal{P'}\setminus B$.
Hence there exists $A_j\in\Gamma_M \ (j\neq1)$
such that $A_j\subseteq\mathcal{P'}\setminus B$, that is, $\mathcal{P'}\setminus B$ is authorized.
For otherwise $A_j\nsubseteq\mathcal{P'}\setminus B$ for any $A_j$ in $\Gamma_M$, then $A_1\cap A_2\cap\cdots\cap A_r\neq\emptyset$, i.e., there exists $P_l$ such that $P_l\in A_1\cap A_2\cap\cdots\cap A_r$. Then we can obtain that $\{P_l\}$ and $\{P_1\cdots P_{l-1}P_{l+1}\cdots P_{n}\}$ are unauthorized sets. This is contrary to the maximal quantum access structure $\Gamma_M$.

By the induction hypothesis, it implies that $r= n-1$. Therefore, we can get that $r+1= n-1+1=n$.
So this proposition is true for $n$ players. This completed the proof.
}

\begin{Cor}\rm
 If $\mathcal{P}$ is a set with $n$ players and $\Gamma_M$ is a maximal quantum access structure containing $r$ minimal authorized sets, then $r>n$.
\end{Cor}

From the proof of Theorem 6, we have proposed a construction method about the minimal maximal quantum access structure. Compared to the maximal quantum access structure, the minimal maximal quantum access structure is more concise and easier to construct. Therefore, we take the access structure with five participants as an example.
Suppose that $\mathcal{P}=\{P_1,P_2,P_3,P_4,P_5\}$ is a set of players, all minimal maximal quantum access structures on $\mathcal{P}$ can be denoted by
\begin{eqnarray*}
  \Gamma_{M_1}^{(m)}&=&\{P_1P_2,P_1P_3,P_1P_4,P_1P_5,P_2P_3P_4P_5\} \\
  \Gamma_{M_2}^{(m)}&=&\{P_1P_2,P_1P_3,P_2P_3P_4,P_1P_4P_5,P_2P_3P_5\}
\end{eqnarray*}

If we want to revoke a player $P_5$ because of some factors, then we only need to change two authorized sets in $\Gamma_{M_1}^{(m)}$ or $\Gamma_{M_2}^{(m)}$, and then we can reconstruct the new minimal maximal access structure.
If we want to join a player $P_6$, then we also only need to change and add two authorized sets in $\Gamma_{M_1}^{(m)}$ or $\Gamma_{M_2}^{(m)}$.
\begin{figure}[htbp]
  \centering
  \includegraphics[width=7cm]{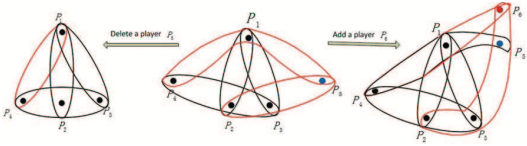}
  \caption{(b)The minimal maximal quantum access structures $\Gamma_{M_2}^{(m)}$; (a)Deleting a player $P_5$ from $\Gamma_{M_2}^{(m)}$; (c)Adding a player $P_6$ to $\Gamma_{M_2}^{(m)}$, where red circle represents the authorized set to change.}
\end{figure}
The FIG.1 shows the minimal maximal quantum access structures $\Gamma_{M_2}^{(m)}$ adds or removes a participant, and we can find that adding or deleting a participant has a minor effect on the minimal authorized sets in the minimal maximal access structure.
Therefore, it is relatively easy to deal with the change of quantum share, which can guarantee the security of secret sharing.

\section{Construction of a General Access Structure}
\subsection{Two Schemes}
In this part, we propose two schemes for general access structure. One is based on decomposition of quantum access structure, and another scheme combines the decomposition of access structure with the minimal maximal quantum access structure.

Scheme I

In the classical secret sharing, there is a perfect secret sharing scheme for general access structure based on the decomposition of access structure. In Sec.II.A, we introduce the decomposition of quantum access structure. Hence we can also propose a quantum secret sharing scheme to realize a general access structure by using the optimal decomposition.

Suppose $\mathcal{P}=\{P_1,P_2,\cdots,P_{n}\}$ is a set of players and $\Gamma$ is a quantum access structure on $\mathcal{P}$. We can find an optimal decomposition $\{\Gamma_1, \Gamma_2,\cdots,\Gamma_l\}$, where each $\Gamma_i$ can be realized by quantum secret sharing protocol. Therefore, we can put these particles held by $P_i$ in the register $R_i$ and then distribute the register $R_i$ to the participant $P_i$ (FIG.2).  Noted that each participant has a register, but the corresponding particles in the different registers are entangled, and the particles in the same register are independent of each other. Any attack will destroy the entanglement between the particles, so that the secret can not be restored.
For different authorized sets, participants can choose different particles and cooperate with others to restore the original secret.
\begin{figure}[htbp]
  \centering
  \includegraphics[width=4cm]{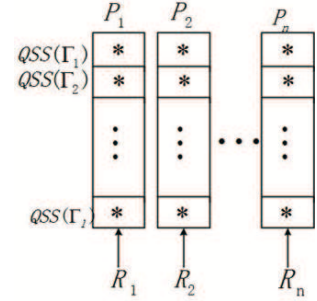}\\
  \caption{Distribution of particles in Scheme I}
\end{figure}

Scheme II

In the following we mainly give the secret sharing to realize a general quantum access structure by the idea of concatenation scheme. Moreover, we also make use of the minimal maximal access structure and the decomposition.

Preparatory phase:
Given a quantum access structure $\Gamma=\{A_1,A_2,\cdots,A_r\}$, where $A_i\in2^\mathcal{P}\ (i=1,2,\cdots,r)$ is the minimal authorized set. By Theorem 3 and Theorem 6, we can obtain the  minimal maximal quantum access structure $\Gamma^{(m)}_M$ from $\Gamma$. Since the fact, we have that for any  maximal quantum access structure there exists a pure quantum secret sharing scheme to realize it.
By Definition 2, it implies that there is a decomposition of the access structure $\Gamma$. Without loss of generality, we may take a decomposition $\Gamma=\Gamma_1\cup\Gamma_2\cup\cdots\cup\Gamma_l\ (l\leq r)$, where $\Gamma_i \ (i=1,2,\cdots,l)$ can be realized by quantum secret sharing protocol. If $l=r$, this decomposition is trivial. Gottesman used this trivial decomposition to design the protocol. If there doesn't exist a positive integer $l'$ such that $l'<l$, this decomposition is optimal. In our paper we will utilize the optimal decomposition to construct QSS in order to save the resources.

Distribution phase:
Due to the the optimal decomposition of $\Gamma$, we get $l$ sub-access structures. According to them, we should take a $((l,2l-1))$ quantum threshold scheme realized in Ref.${[4]}$.

(i) Distribute $l$ shares of $((l,2l-1))$ quantum threshold scheme to $\Gamma_1,\Gamma_2,\cdots,\Gamma_l$
, respectively. Without loss of generality, Share $i$, for $i=1,2,\cdots,l$, is mapping to $\Gamma_i$ as the secret. For each $\Gamma_i$, it can be realized by quantum secret sharing scheme.

(ii) Distribute the remaining shares of $((l,2l-1))$ scheme to a minimal maximal quantum access structure $\Gamma^{(m)}_M$. Since the fact (i), then there exists a pure state scheme to realize $\Gamma^{(m)}_M$.

Reconstruction phase:
We will analyze the access structure of this concatenation scheme and verify the real authorized sets. Only these players in the real authorized sets can cooperatively obtain the original information.

(i) Suppose that a set $A\in\Gamma$ containing certain $A_i$, i.e., $A_i\subseteq A$. We can know that the set $A$ is also an authorized set of $\Gamma^{(m)}_M$ since $\Gamma\subseteq\Gamma_M^{(m)}$. For the authorized set $A$, we can reconstruct the $l$ shares of the $((l,2l-1))$ scheme, where $l-1$ shares are from $\Gamma^{(m)}_M$ and the only one from $\Gamma_i$. Hence $A$ is also an authorized set of the concatenation scheme.

(ii) Suppose that there is a set $B$ such that $A_i\nsubseteq B$ for any $A_i\in\Gamma$. Thus $B$ is an unauthorized set of $\Gamma$. If $B$ is an authorized set of $\Gamma^{(m)}_M$, then we can only reconstruct the $l-1$ shares from $\Gamma^{(m)}_M$. Hence $B$ is also an unauthorized set of the concatenation scheme.

From the above (i) and (ii) we can know that the access structure of the concatenation scheme is exactly $\Gamma$, that is, only the sets of $\Gamma$ are authorized ones that can restore the original secret.
\subsection{Comparision}
In this section, we discuss the comparison of Scheme I and Scheme II. In addition, we also compare our scheme II and Gottesman's construction by example.

In Scheme I, we make use of the decomposition of quantum access structure and know that it is easy to find an optimal decomposition. Hence the advantage of this scheme is to achieve a general quantum access structure.
In this scheme each participant holds many particles, but register storage capacity is limited. If each participant has too many information shares, it may lead that the register capacity is insufficient. In addition, each participant directly grasped a large amount of information shares about the original secret. If the scheme was attacked by the participants conspiracy, it is easy to cause the leakage of the original information.

In Scheme II, we use the idea of concatenation scheme and combine the minimal maximal access structure and the decomposition.
Compared to Scheme I, the original secret in Scheme II will be divided into some secret shares, and we treat each share as a secret to each sub-access structure. So it can ensure that the participants do not have directly access to the secret share and reduce the chance to leakage of information. In this scheme, participants first cooperate to recover the secret shares and then cooperate to restore the original secret. Hence Scheme II is more secure and we give an example.

\begin{Ex} \rm
In Example 1, we have given the quantum access structure $\Gamma=\{P_1P_2,P_1P_4P_5,P_2P_3P_5,P_2P_3P_4\}$. For this access structure, we can add some sets to obtain a minimal maximal quantum access structure $\Gamma_M^{(m)}=\{P_1P_2,P_1P_3,
P_1P_4P_5,P_2P_3P_5,P_2P_3P_4\}$. Moreover, we can find an optimal decomposition of $\Gamma$. It is denoted by $\Gamma=\Gamma_1\cup\Gamma_2$, where $\Gamma_1=\{P_1P_2,P_1P_4P_5\}$ and $\Gamma_2=\{P_2P_3P_5,P_2P_3P_4\}$. In Ref.${[33]}$, there exists a generalized quantum secret sharing scheme to realize $\Gamma_i\ (i=1,2)$ (GQSS).
Hence we can consider the $((2,3))$ quantum threshold scheme. The three rows represent shares of a $((2,3))$ scheme, so authorized sets on any two rows suffice to reconstruct the secret.
\begin{eqnarray*}
 ((2,3)) scheme \left\{
   \begin{array}{ll}
   GQSS:& \Gamma_1=\{P_1P_2,P_1P_4P_5\}\\
   GQSS:& \Gamma_2=\{P_2P_3P_5,P_2P_3P_4\}\\
   \Gamma_M^{(m)}&
   \end{array}\right.
  \end{eqnarray*}
The first two rows are threshold schemes. $\Gamma^{(m)}_M$ is a minimal maximal quantum access structure containing $\Gamma_1$ and $\Gamma_2$. It is easy to verify that the set $P_1P_3$ is unauthorized.

For our construction method, we firstly divide the quantum access structure $\Gamma$ into two parts,  $\Gamma_1$ and $\Gamma_2$. This is an optimal decomposition. According to the optimal decomposition, we adopt $((2,3))$ quantum threshold scheme realizing the access structure $\Gamma$. In Ref.${[5]}$, Gottesman gave a trivial decomposition of $\Gamma$, so he would use the $((4,7))$ scheme to realize the same access structure (see below). Obviously, his scheme is more cumbersome and uses more quantum shares than ours.
\begin{eqnarray*}
 ((4,7)) scheme \left\{
   \begin{array}{ll}
   ((2,2)):& \{P_1P_2\}\\
   ((3,3)):& \{P_1P_4P_5\}\\
   ((3,3)):& \{P_2P_3P_5\}\\
   ((3,3)):& \{P_2P_3P_4\}\\
   \Gamma_M&
   \end{array}\right.
  \end{eqnarray*}
\end{Ex}

Compared to Gottesman's construction, we utilize a minimal maximal quantum access structure instead of a
maximal one. On one hand, each maximal quantum access structure is included by the minimal maximal one. On the other hand,  because the minimal maximal quantum access structure reduces the number of the minimal authorized sets, we will greatly reduce the number of tests in the process of verifying the authorized set. Furthermore, the efficiency of the scheme will be greatly improved. In TABLE I, we give a comparison between them containing five or six participants.
Moreover, with the increase of the number of participants, the construction of the maximal access structure is difficult. However, the minimal maximal access structure is easily obtained by our method in Theorem 6. In addition, it is easy to find that our scheme based on the optimal decomposition is more convenient and save more quantum resources. Hence the optimal decomposition of quantum access structure is also valid for the construction of secret sharing schemes.
\begin{table*}[htbp]
\caption{Comparison of  verifying the real authorized set with five or six participants}
\begin{tabular}{c|c|c}
        \hline
        {the Number of participants}&{Quantum access structure ($\Gamma_M,\Gamma^{(m)}_M$)} & {Verification times} \\
        \hline
        \multirow{3}{0.5cm}{5} & {$\Gamma_M=\{P_1P_2P_3,P_1P_2P_4,P_1P_2P_5,P_1P_3P_4,P_1P_3P_5,$} & {10}\\
        & {$ P_1P_4P_5,P_2P_3P_4,P_2P_3P_5,P_2P_4P_5,P_3P_4P_5\}$} & \\\cline{2-3}
                &{$\Gamma^{(m)}_M=\{P_1P_2,P_1P_3,P_1P_4P_5,P_2P_3P_4,P_2P_3P_5\}$} & {5}\\\cline{2-3}
        \hline
        \multirow{3}{0.5cm}{6} & {$\Gamma_M=\{P_1P_2,P_1P_3P_4,P_1P_3P_5,P_1P_3P_6,P_1P_4P_5,P_1P_4P_6,P_1P_5P_6,$}&{11}\\
        &{$P_2P_3P_5P_6,P_2P_4P_5P_6,P_2P_3P_4P_5,P_2P_3P_4P_6\}.$} &\\\cline{2-3}
                &{$\Gamma^{(m)}_M=\{P_1P_2,P_1P_3,P_1P_4,P_1P_5P_6,P_2P_3P_4P_5,P_2P_3P_4P_6\}$}& {6}\\\cline{2-3}
        \hline
        \end{tabular}
\end{table*}

\section{Conclusions and Discussion  }
In this work firstly we proposed a definition about decomposition of the quantum access structure. Secondly,
we formally defined a maximal quantum access structure. After analysing it we also presented a minimal maximal quantum access structure. Next, we gave a sufficient and necessary condition to determine the minimal maximal quantum access structure and gave other conclusions about it. We discussed the relationship between the number of the minimal authorized sets in minimal maximal access structure and that of participants. Finally, we gave the application about a decomposition of the quantum access structure and a minimal maximal access structure in secret sharing. Then
we proposed two quantum secret sharing schemes to realize a general access structure.
Our scheme II was based on the method of concatenation and decomposition of an access structure. Compared to the existing scheme,
our scheme II can save quantum resources and reduce the cost.

In addition, for QSS many factors may lead that the access structure in the secret sharing is changed, such as the security requirements and changing to the participants in the attack. Therefore, a dynamic secret sharing scheme has very important research value ${[34-36]}$. If a participant in the system is suspected because it may be compromised, we can change the access structure to reduce the role of this member in the reconstruction phase. Hence it can continue to maintain security of the whole system. Compared with the normal secret sharing scheme, dynamic secret sharing has higher security and greater flexibility in application.
If the dynamic scheme uses a minimal maximal access structure, then this process will be relatively easy to add or delete a participant. Therefore, it is simple to deal with the change of quantum share, which can guarantee the security of secret sharing. This is an interesting question, and we can further study how to give a specific dynamic secret sharing scheme to realize the minimal maximal quantum access structure.

\section*{ACKNOWLEDGEMENT}
We want to express our gratitude to anonymous referees
for their valuable and constructive comments. This work was sponsored by the National Natural Science Foundation of China under Grant No.61373150 and No.61602291, and Industrial Research and Development Project of Science and Technology of Shaanxi Province under Grant No.2013k0611.

\end{document}